# Bcl-2 inhibits apoptosis by increasing the time-to-death and intrinsic cell-to-cell variations in the mitochondrial pathway of cell death

Joanna Skommer · Tom Brittain · Subhadip Raychaudhuri

**Abstract** BH3 mimetics have been proposed as new anticancer therapeutics. They target anti-apoptotic Bcl-2 proteins, up-regulation of which has been implicated in the resistance of many cancer cells, particularly leukemia and lymphoma cells, to apoptosis. Using probabilistic computational modeling of the mitochondrial pathway of apoptosis, verified by single-cell experimental observations, we develop a model of Bcl-2 inhibition of apoptosis. Our results clarify how Bcl-2 imparts its anti-apoptotic role by increasing the time-to-death and cell-to-cell variability. We also show that although the commitment to death is highly impacted by differences in protein levels at the time of stimulation, inherent stochastic fluctuations in apoptotic signaling are sufficient to induce cell-to-cell variability and to allow single cells to escape death. This study suggests that intrinsic cell-to-cell stochastic variability in apoptotic signaling is sufficient to cause fractional killing of cancer cells after exposure to BH3 mimetics. This is an unanticipated facet of cancer chemoresistance.

J. Skommer · T. Brittain
School of Biological Sciences, University of Auckland, Thomas Bld., 3a Symonds Street, Auckland 1142, New Zealand
e-mail: J.Skommer@auckland.ac.nz

T. Brittain
e-mail: T.Brittain@auckland.ac.nz

S. Raychaudhuri (✉)
Department of Biomedical Engineering, University of California, 2521 Genome and Biomedical Sciences Bld., 451 Health Sciences Drive, Davis, CA 95616-5294, USA
e-mail: raychaudhuri@ucdavis.edu

**Keywords** Systems biology · Bcl-2 · Apoptosis · Stochastic variability · Cancer

## Introduction

Apoptosis (programmed cell death) is an evolutionarily conserved form of cell death which allows the removal of damaged or superfluous cells in order to maintain tissue homeostasis. It can be mediated by two pathways, the death receptor pathway and the mitochondrial (intrinsic) pathway. The latter is finely regulated up-stream of mitochondria by the concerted action of many molecules, including a family of Bcl-2 proteins which consists of both pro-apoptotic (BH3-only proteins and multidomain proteins) and anti-apoptotic (Bcl-2-like proteins) members [1]. Interactions between these proteins control the process of cytochrome *c* release from the mitochondrion, modulating the sensitivity to cell death signals [1]. BH3-only proteins belonging to this family have been suggested to induce cell death by restraining the anti-apoptotic Bcl-2 proteins and/or directly activating multidomain pro-apoptotic Bax/Bak proteins [1, 2]. Interestingly, some of the oncogenic events, such as genomic instability, oncogene activation or loss of adhesion, can directly activate BH3-only proteins and either induce or sensitize cells to apoptosis [3]. To modify these death signals, cancer cells often increase the levels of anti-apoptotic factors, such as Bcl-2, becoming dependent on this anti-apoptotic protein [4, 5]. In such cells BH3 mimetics can induce apoptosis in a single agent treatment scenario, by displacing the bound BH3-only proteins and allowing activation of Bax/Bak [3].

Even though Bcl-2 inhibition of apoptosis has been extensively studied in the past (reviewed in [6–8]), how varying levels of a single form of anti-apoptotic Bcl-2-like protein translate into clear cell fate outcomes needs to be elucidated at the molecular level. This becomes even more important as recent mathematical and experimental studies suggest that non-genetic cell-to-cell variability is central to the signaling in the intrinsic pathway of apoptosis and seems to explain an unusually slow cell death through this pathway [9, 10]. A crucial feature of such variability is that a set of genetically and epigenetically identical cells can respond to an apoptotic stimulus in a very different manner [9–13].

Here, using a combination of probabilistic computational modeling, flow cytometry and single-cell microscopy data, we study the concentration-dependent variability in Bcl-2 inhibition of apoptosis. The experimentally-verified computational model of apoptosis reproduces the behavior of a heterogeneous population of cells treated with a BH3 mimetic and shows how varying levels of Bcl-2 regulate the time-to-death ($T_d$) and cell-to-cell variability in caspase activation and cell death. Increased expression of Bcl-2 slows down apoptosis, increasing also the cell-to-cell variability in time to apoptotic death. Thus, a cell with particularly long time-to-death might escape apoptotic death and initiate tumor formation. For cancer cells our results show that treatment of a hypothetical population of identical cancer cells still leads to fractional cell killing, which results solely from the intrinsic stochastic variability in chemical reactions. This challenges the concept that cell-to-cell variability in execution of apoptosis arises only due to genetic or epigenetic differences, or varying protein states and concentrations, and clearly affects our expectations as to the efficacy of apoptosis-targeted anticancer therapies.

**Materials and methods**

Cell culture and reagents

Jurkat leukemia T-cells were maintained in RPMI 1640 (Invitrogen) with L-glutamine, 1% penicillin/streptomycin and 10% fetal bovine serum (FBS; Invitrogen) at 37°C in humidified 95% air, 5% $CO_2$. Neuroblastoma SH-SY5Y cells were maintained in advanced DMEM/F12 (1/1) (Invitrogen), whereas HEK293 cells in DMEM (Invitrogen), supplemented as above. To induce apoptosis, cells were treated with a small-molecule BH3 mimetic HA14-1 (Alexis Biochemicals). HA14-1 led to mitochondrial membrane depolarization in all dying cells, as assessed by flow cytometry combining 7-AAD and TMRM (below).

Analysis of apoptotic cell death

To assess cell viability, $0.3 \times 10^6$ Jurkat T cells or $0.1 \times 10^6$ SH-SY5Y cells were plated on 24-well plates, and treated as indicated. At the end of the experiment, live and dead cells were collected, washed with PBS, and stained for 20 min at RT with Annexin V-PE (Invitrogen) and 7-AAD (Invitrogen; 1 μg/sample) in the Annexin V binding buffer (Invitrogen). The cells were analyzed immediately on FACS Calibur (BD). Cells were gated based on forward scatter (FSC) and side scatter (SSC) to exclude cell debris, and next analyzed based on Annexin V-PE fluorescence and 7-AAD fluorescence using CellQuest (BD). Plots were generated using WinMDI 2.8.

Colony forming assay

SH-SY5Y cells were plated at $5 \times 10^4$ cells/well in 12-well plates. The next day the cells were treated with DMSO or HA14-1 (12.5 μM) for 48 h. Cells were then washed, and cultured in fresh medium. Cells were cultured over 10–12 days in DMEM/F12 supplemented with 10% FBS, penicillin/streptomycin, and glutamine. Medium was replaced every 3 days. Cell colonies were stained with Giemsa, fixed in 10% methanol, and cells photographed, and then counted using ImageJ software.

Analysis of caspase 9 activity and multiparameter (FLICA/TMRM/7-AAD) flow cytometry assay

Activation of caspase 9 was examined by a combination of fluorescently labeled inhibitor of caspase (FLICA) FAM-LEHD-FMK (Calbiochem) and 7-AAD (Invitrogen; 1 μg/sample) as a probe of early plasma membrane permeability. We have previously determined that the maximal fluorescence of FLICA reagent corresponds to the complete activation of caspase 9 [14]. Cells ($0.15 \times 10^6$ per well) were cultured for the time indicated with or without the indicated doses of HA14-1, harvested, and incubated with FLICA for 45 min at 37°C under 5% $CO_2$ in darkness. Then, cells were washed three times, stained with 7-AAD, and immediately analyzed on a FACS Calibur (BD). For the three-color analysis, FLICA-stained cells were also labeled with TMRM (150 nM, Invitrogen) for 30 min at 37°C under 5% $CO_2$ in darkness, followed by staining with 7-AAD and immediate analysis. Cells were gated based on forward scatter (FSC) and side scatter (SSC) to exclude cell debris, and based on FSC versus 7-AAD to exclude cells with plasma membrane permeability. Next cells were analyzed based on FLICA fluorescence alone or in combination with analysis of TMRM fluorescence. Plots were generated using WinMDA.

Time-lapse epifluorescent microscopy and digital imaging

For analysis of cell-to-cell variability in time to death, SH-SY5Y cells were treated in the presence of propidium iodide (PI; 0.5 μg/ml) to monitor appearance of dying cells with loss of plasma membrane integrity. Image acquisition was performed every 15 min. The presence of PI in tissue culture media has been previously reported not to affect the outcome of cytotoxicity assays [15]. To analyze cell-to-cell variability in mitochondrial membrane depolarization, SH-SY5Y cells and HEK293 cells were first equilibrated with TMRM (200 nM; Invitrogen) in DMEM/F12 medium and DMEM medium, respectively, for at least 2 h at 37°C. This approach to study of MOMP has the advantage of avoiding the need for plasmid transfections and its associated artifacts. After equilibration and baseline imaging, cells were treated on stage with varying doses of HA14-1. Image acquisition was performed for 4 h every 2 min. Time-lapse imaging was performed on a Nikon Ti-E inverted microscope (Nikon Corp., Kawasaki, Japan) equipped with a ProScan H117 motorized XY stage and NanoScan NZ250 piezo Z stage (both Prior Scientific Instruments Ltd, Cambridge, UK), and an Andor iXon DU-885 EM-CCD camera. The imaging setup was controlled by Andor iQ v.1.9 software (Andor Technology Ltd, Belfast, UK). A Semrock Cy3 4040B filter cube (excitation 531/40 nm, dichroic mirror 562 nm, emission 593/40 nm; Semrock, Rochester, NY, USA) was used for epifluorescence. Cells were maintained at 37°C with supplemental $CO_2$ using the incubator cabinet (Clear State Solutions Pty Ltd, Melbourne, Australia).

Image processing

After background subtraction, the cellular TMRM fluorescence intensity was calculated for each cell. At least ten cells were analyzed for each treatment, with quantification of 120 images per cell. Images were processed using ImageJ. A difference of more then the SD below the initial baseline, which did not recover to baseline values, was defined as onset time of $\Delta\psi_m$ loss. In contrast to HA14-1, the mitochondrial uncoupler FCCP led to simultaneous loss of $\Delta\psi_m$ in all the cells (Fig. S1), confirming that cell-to-cell variability in time to $\Delta\psi_m$ loss is not the intrinsic feature related to TMRM labeling or functioning of the mitochondrial electron chain.

Computational systems analysis

We used a Monte Carlo stochastic simulation model of apoptotic signaling reactions that explicitly simulates diffusion and reaction moves at the level of individual molecules [11, 12]. At the beginning of the simulation all signaling molecules are distributed randomly and uniformly in the cell volume simulated by a three dimensional cubic lattice. Cytochrome c molecules were confined within a fixed mitochondrial region inside the cell volume. Once the concentration of active Bax dimers reaches a pre-assigned threshold value (∼0.017 μm), cytochrome c is released from the mitochondria into the cytosol in an all-or-none manner. Recent experiments indicate that only when Bax activation reaches a threshold, formation of lipidic pores in mitochondrial membrane leads to release of cytochrome c [16], which justifies our model assumptions regarding Bax dimerization-induced cytochrome c release. Other mechanisms such as mitochondrial membrane depolarization due to cumulative effects of Bax activation would result in a qualitatively similar signaling response, however, with a slightly modified time-scale of apoptosis activation.

In our simulations, truncated Bid (tBid) initiates the apoptotic signaling. At each attempt of Monte Carlo move, a molecule is picked randomly, and either diffusion or a reaction move is performed with pre-assigned probability values. We assume mutual physical exclusion of signaling molecules while carrying out the diffusion move. Thus, for example, while performing a Monte Carlo move of binding reactions two molecules that are on adjacent sites are allowed to react. Explicit spatial simulation allows us to model spatial localization such as: (a) translocation of active Bax dimers onto mitochondrial membranes or (b) clustering of multiple cytochrome c and APAF-1 molecules in an apoptosome assembly. In addition, explicit spatial simulation allows us to assign differential diffusion probability $P_{diff}$ for different types of free and complex molecules. For example, multi-molecular apoptosome complexes are assumed immobile in a crowded cellular environment.

We defined separate functions for some general form of reactions, such as (i) two molecules can bind in their free binding pockets to form a complex with a probability $P_{on}$, (ii) a molecule can dissociate from a multi-molecular complex with a probability $P_{off}$, (iii) a signaling molecule, while dissociating from a complex, can undergo conformational change (or cleavage) to an active form with probability $P_{cat}$. At the beginning of the simulation all the molecules were placed in a random and uniform manner. At each Monte Carlo attempt a randomly selected molecule is allowed to undergo either diffusion or a reaction move. In our simulations, we kept track of the number of molecules of each signaling species over a period of time.

The proteins with similar biochemical activities were represented by single species in our simulations. For instance, caspase 7, which is functionally similar to caspase 3, as well as proteins functionally redundant to XIAP (i.e., c-IAP1, c-IAP2 and NAIP), were not explicitly included in our model. Similarly, Bcl-2 represented all anti-apoptotic Bcl-2 family members, whereas Bax represented pore-forming proteins. Therefore, the protein concentrations varied in our simulations represent sums of functionally redundant protein concentrations. We studied the effect of physiologically relevant protein variations in apoptosis signaling by sampling Gaussian probability distributions. The *physiologically relevant* protein variations refer to cell-to-cell variability in protein concentrations that may exist at the time of application of an apoptotic stimulus and originated from noisy gene expression [17–20]. The average of the distribution for a specific protein was assumed to be the known average concentration for that particular signaling species. We used a coefficient of variation (CV, variance/mean) of 0.25 in the Gaussian distributions for all signaling proteins [10, 20]. At the beginning of our simulation we randomly sampled from respective Gaussian distributions and assigned different concentrations of proteins to individual simulation runs (single cells in a population). We did not allow further protein additions (de novo synthesis) or degradations during the course of the simulation.

Estimates for model parameters, such as rates, initial protein concentrations, and IC50 for HA14-1 binding to Bcl-2 were obtained from the literature [9, 21–25]. An earlier developed parameter-mapping scheme was used to estimate the values of probabilistic constants, which are directly used in our simulations, from experimentally known values of physical variables [11–13]. In this approach, the time-scale of simulation emerges naturally from our Monte Carlo simulations rendering a direct comparison with experimentally observed time-scales possible.

## Results

The complex nature of protein signaling networks, such as apoptotic pathways, with multiple variables acting at the same time, can be often best studied using a systems level analysis. Therefore, using systems level information of the apoptotic signaling reactions, we have developed a quantitative model of the mitochondrial pathway of apoptosis (Fig. 1a). This approach is based on a probabilistic method

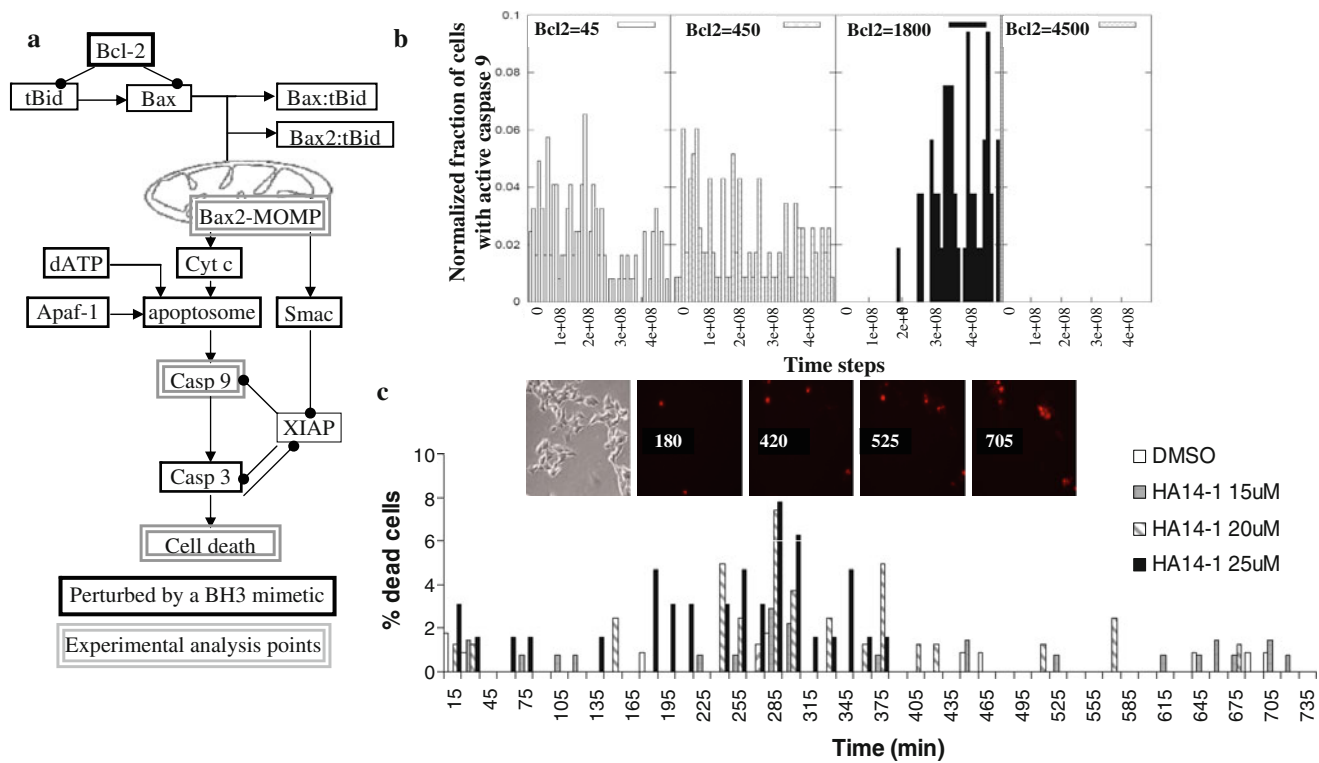

**Fig. 1** Probabilistic computational model of the intrinsic pathway of apoptosis reproduces cell-to-cell variability in time-to-death. **a** Schematic of the mitochondrial pathway network modeled in this study. *MOMP* mitochondrial outer membrane permeabilisation, *tBid* truncated Bid. **b** Cell-to-cell variability and time to activation of caspase 9 depend on Bcl-2 concentration. *Insets* indicate the number of Bcl-2 molecules, corresponding to ∼0.075 µM (45 molecules), ∼0.75 µM (450 molecules), ∼3 µM (1800 molecules), and ∼7.5 µM (4500 molecules). Time is measured in Monte-Carlo (MC) simulation steps. 1 MC step = $10^{-4}$ s, hence time-scale shown $5 \times 10^8$ MC steps ∼15 h. **c** *Upper panel* time-lapse images of SH-SY5Y cells treated with 10 µM HA14-1. Bright field image was taken at time 0. *Inset numbers*, time (min) after treatment. Dead cells become permeable to propidium iodide (PI). *Lower panel* the non-cumulative frequency distribution in time-to-death for SH-SY5Y cells treated with HA14-1 at concentrations indicated, as determined by live-cell microscopy (n > 70 for each condition)

in which the reactivity of all the signaling molecules follows a stochastic, rather than deterministic behavior, and probes the induction of cell death at a single cell level [11].

We simulated the induction of the mitochondrial pathway of apoptosis by allowing the direct tBid–Bax reaction to generate pro-apoptotic Bax dimers [2]. The input signal used in the model was varying concentration of Bcl-2 available to inhibit tBid and Bax molecules, which experimentally was recapitulated using a small molecule BH3 mimetic HA14-1 [21], previously shown to bind and titrate down Bcl-2, relieving its inhibitory action on Bax (indirect activation model) [26]. In our computational model cytochrome $c$ was released in an all-or-none manner once a threshold number of Bax dimers (denoted as Bax2) are formed. This is in line with recent experimental studies suggesting that the release of cytochrome $c$ occurs rapidly ($\sim 5$ min) once it is initiated [27, 28]. Our model assumes also that cytochrome $c$, once released into the cytosol, binds with Apaf-1 and dATP to form the apoptosome, with subsequent activation of caspase 9 and caspase 3. The release of Smac has been modeled to inhibit anti-apoptotic XIAP proteins. XIAP can inhibit procaspase-9 as well as activated molecules of caspase-9 and caspase-3 (Fig. 1a). Active caspase 3 decreases the concentration of anti-apoptotic XIAP, creating an effective positive feedback loop in the pathway (Fig. 1a). The proteins with similar biochemical activities were represented by single species in our simulations. For example, Bax represents all pro-apoptotic multi-domain Bcl-2 proteins (i.e., both Bax and Bak), whereas Bcl-2 represents all anti-apoptotic Bcl-2 proteins (e.g., Bcl-2, Bcl-X$_l$, Mcl-1, etc.).

Using the simplified model of the mitochondrial pathway of apoptosis we seek to understand how varying levels of the anti-apoptotic Bcl-2 protein affect the cells commitment to apoptosis and cell-to-cell variability in time-to-death ($T_d$). In order to make the computational modeling amenable, the initial approach was to keep the coefficient of variation (CV) of all protein concentrations equal to 0. In a clonal cell population, which was used for experimental validation of the model, the concentrations of proteins regulating apoptosis are expected to vary, with CV ranging from 0.21 to 0.28 [10]. Despite this approximation, the time scale of apoptosis that emerges naturally from our stochastic simulations show slow (approx. hours) activation that becomes prolonged with increasing Bcl-2 concentrations (Fig. 1b, Fig. S2), as also observed in our experimental studies (Fig. 1c). With single-cell time lapse microscopy we show also that cell-to-cell variability in $T_d$ depends on the dose of HA14-1, and is particularly heterogeneous at low doses of HA14-1 (Fig. 1c). This is recapitulated well by our simplified model of apoptosis showing an all-or-none activation of caspase 9 (and caspase 3) with increasing cell-to-cell variability as the concentration of Bcl-2 is increased (Fig. 1b, Fig. S2). Importantly, cell-to-cell variability in $T_d$ cannot be completely abolished, and is not sensitive to reduction of Bcl-2 levels below a certain threshold (here $\sim 0.75$ μM) (Fig. 1b). Intermediate Bcl-2 level (here $\sim 3$ μM) allows induction of apoptosis but with a long time lag before caspases are activated and with very large cell-to-cell variation. This explains how a rare cell might escape apoptotic death and, if it has tumor-promoting features, initiate or accelerate tumor growth. For very high Bcl-2 levels (here $\sim 7.5$ μm), as possibly found only in some cancer cells, apoptotic activation does not take place within the time-scale of our simulations ($\sim 15$ h). This seems to explain unusual apoptosis resistance of cancer cells that have high over-expression of Bcl-2.

We also calculated the probability of caspase activation in order to characterize cell-to-cell stochastic fluctuations with the decreasing concentration of Bcl-2 protein. The combined effects of (a) all-or-none activation of caspase 9 and 3 at the single cell level, and (b) cell-to-cell fluctuations lead to bi-modal probability distributions for activated caspase 9 (Fig. 2a) and caspase 3 (Fig. S3). Such bi-modal probability distributions can be used to estimate the ratio variance/average, which is a quantitative estimate of cell-to-cell variability in apoptosis activation. If we approximate the probability distributions as perfect bi-modal curves, the ratio variance/average for caspase 3 activity is found to be C3[1-f(Bcl-2,t)], where f(Bcl-2,t) denotes the fraction of cells in which caspase 3 has undergone complete activation at a given time $t$. Clearly, f(Bcl-2,t) decreases with increasing Bcl-2 concentration thus indicating an increase in cell-to-cell variability due to decreased Bcl-2 inhibition. We also determine the distribution of caspase 9 activity in HA14-1-treated immortalized Jurkat T-cell leukemia cells, using a multiparameter flow cytometry assay based on FAM-LEHD-FMK (caspase 9 FLICA) reagent which binds to the activated caspase 9, TMRM which labels energized mitochondria, and plasma membrane integrity marker 7-AAD (Fig. 2b). Caspase 9 activity increases slightly in HA14-1-treated Jurkat T cells that have lost the mitochondrial transmembrane potential ($\Delta\psi_m$) (Fig. 2b), as measured when cells are gated for negative labeling by 7-AAD and for lack of changes in forward versus side scatter. Another population of cells exhibited loss of $\Delta\psi_m$ and over tenfold increase in FLICA fluorescence (Fig. 2b). Importantly, the intermediate events in terms of caspase activity are infrequent (<2%) (Fig. 2b), in line with the concept of bistability. We also analyze caspase 9 activation in Jurkat T cells treated with increasing doses of HA14-1. With the increase in the dose of HA14-1, the population of cells with activated caspase 9 becomes more abundant, with early caspase 9 activation represented by a slight increase in FLICA fluorescence, and complete caspase 9 activation represented by over tenfold increase in FLICA fluorescence, and with very sparse (<5%) intermediate events (Fig. 2c). We also observed such a bimodal distribution of caspase activity, with infrequent intermediate events between the initial and full caspase 9 activation, in SH-SY5Y cells (Fig. S4). Finally, for a fixed Bcl-2 concentration, the simplified model of the intrinsic pathway of apoptosis predicts well the kinetics of caspase 9 activation observed in Jurkat T cells treated with HA14-1 (Fig. 2d).

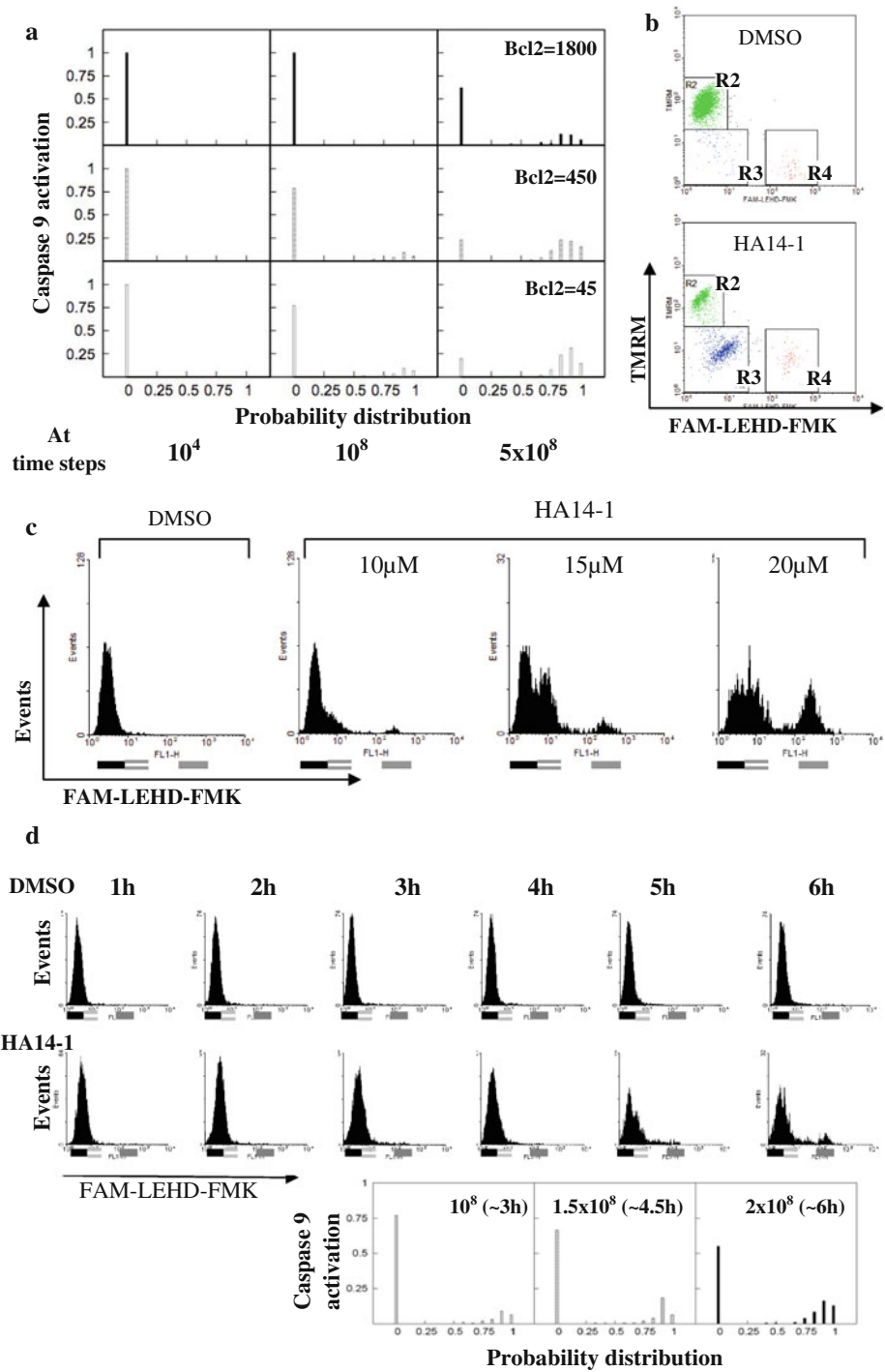

Next we asked whether variability upstream of mitochondria can explain cell-to-cell fluctuations in $T_d$. In our computational model, formation of a critical number of Bax2 dimers leads to immediate release of mitochondrial

◂**Fig. 2** Probabilistic computational model of the intrinsic pathway of apoptosis predicts the kinetics and bi-modal distribution of caspase activation. **a** Probability distribution of caspase 9 activation calculated from single cell activation data of caspase 9. *Insets* indicate the number of Bcl-2 molecules, corresponding to ∼0.075 μM (45 molecules), ∼0.75 μM (450 molecules), and ∼3 μM (1800 molecules). Time is measured in Monte-Carlo (MC) steps. 1 MC step = $10^{-4}$ s. **b** Analysis of early caspase 9 activation in cells treated with HA14-1 confirms rapid caspase 9 activation without intermediate events. Jurkat T cells were treated with HA14-1 (15 μM) for 6 h, labeled with FAM-LEHD-FMK (caspase 9 FLICA), TMRM and 7-AAD, and analyzed by flow cytometry. Late apoptotic events were excluded based on changes in 7-AAD versus forward scatter plots. Region R2, cells with energized mitochondria and lack of caspase 9 activation; region R3, cells with loss of $\Delta\psi_m$ and no or initial caspase 9 activation; region R4, cells with loss of $\Delta\psi_m$ and full caspase 9 activation. Note that there are nearly no intermediate events between the regions R3 and R4. **c** Jurkat T cells were treated for 6 h with HA14-1 at concentrations indicated, and analyzed as above. Note a characteristic shoulder on FLICA fluorescence plots (marked below the plots with ▭), probably indicative of early caspase 9 activation, as well as small number of cells having intermediate caspase 9 activation. Cells with no caspase activation are marked with ▬, and cells with full caspase activation are marked with ▬. **d** *Upper panel* the kinetics of caspase activation, analyzed using caspase 9 FLICA in 7-AAD negative Jurkat T cells treated with DMSO or 10 μM HA14-1; *lower panel* probability distribution of caspase 9 activation calculated from single cell activation data of caspase 9, for 45 molecules of Bcl-2 (∼0.075 μM) cytochrome *c*. We studied the activation of Bax2 complexes as we varied the concentration of Bcl-2 in our simulations. Increasing the Bcl-2 concentration had a strong impact on Bax activation, which slowed down in a noticeable manner with significantly increased cell-to-cell variability as measured by time-to-activation of Bax2 (Fig. 3a). This is also seen from analyzing cell-to-cell variability in time to onset of the loss of TMRM fluorescence in SH-SY5Y cells. The onset of Bax translocation to mitochondria and formation of Bax oligomers usually does not differ significantly from the onset of the loss of mitochondrial transmembrane potential ($\Delta\psi_m$), which can be detected by decreased fluorescence of TMRM [29]. Previous studies have also indicated that the initial loss of TMRM fluorescence coincides with MOMP and the release of cytochrome *c* [30]. Here we observe that HA14-1-induced loss of TMRM fluorescence (Fig. 3b) occurs with large cell-to-cell variability which decreases with the increase in the dose of HA14-1 (Fig. 3c, Fig. S5). Thus, our computational model reproduces qualitatively the behavior of cell population exposed to the BH3 mimetic. Cell-to-cell variability in Bax activation, as observed in our simulations, is non-genetic and arises due to inherent stochasticity of signaling reactions. Such stochastic fluctuations occur in the presence of a large number of Bax and Bcl-2 molecules, and actually increases as the Bcl-2 concentration is increased [17–19]. Even though a large number of Bax and Bcl-2 molecules are initially present in cells, high Bcl-2 concentration can lead to activation of only a few Bax molecules and thus can serve as a source of signaling noise in apoptosis. Cell-to-cell variability in Bax2 activation is readily translated to cell-to-cell variability in the release of cytochrome *c*. There is a second amplification in cell-to-cell stochastic fluctuations post cytochrome *c* release (compare 3a with 1b, and Fig. 3c with 1c), due to low probability of apoptosome formation, but the impact of Bcl-2 on the cell-to-cell variability remains preserved.

We also analyzed the kinetics of mitochondrial membrane depolarization ($\Delta\psi_m$) and activation of caspase 9 using TMRM/FLICA/7-AAD assay and flow cytometry. Mitochondria are affected first by the treatment with HA14-1, which initially does not induce caspase activation (Fig. 3d). At a cell population level, loss of $\Delta\psi_m$ occurs gradually rather then in an all-or-nothing fashion (Fig. 3d, note the lack of clear separation between cell subpopulations on *y* axis), and pro-longed treatment (≥4 h) leads to gradual accumulation of cells with maximum loss of $\Delta\psi_m$. Notably, caspase 9 activation is initiated with some delay and in the population of cells with the lowest fluorescence of TMRM (Fig. 3d). Moreover, at any time point, the population of cells with activated caspase 9 was proportionally low within live/early apoptotic cells, as they very rapidly proceeded to cell death characterized by the plasma membrane permeability to 7-AAD (Supplementary Table 1). These data suggest that upon treatment with HA14-1 mitochondria act as the integrator of death signal, as suggested previously [10, 13, 31], and confirms that cell-to-cell variability in time-to-MOMP contributes significantly to the observed cell-to-cell variability in $T_d$.

We and others have repeatedly observed that exposure to BH3 mimetics elicits a non-uniform reaction in a clonal cell population [14], with cells responding at different time points and some not responding at all (Fig. 1c, Fig. S6). Genetic and epigenetic differences as well as *subtle* variations in protein concentrations and states contribute significantly to the experimentally observed bifurcation in cell fate [10, 20]. The *subtle* cell-to-cell variability in protein concentrations can arise due to noisy gene expressions even within a set of genetically identical cells [17–20]. Here we test how cell-to-cell variability in time-to-death ($T_d$) is affected by the inherent stochastic variability in apoptotic signaling as compared with *subtle* physiologically-relevant variations in protein concentrations. To this aim we include in our model the explicit inhibition of Bcl-2 by HA14-1. We then test cell-to-cell variability in $T_d$ as we vary the concentration of HA14-1 in the model and sample the Bcl-2 concentration from a Gaussian distribution with coefficient of variation (CV) of 0.25 [10, 20] or 0 (hypothetical population of identical cells), and mean concentration of 3.75 and 7.5 μM (Fig. 4, Fig. S7). We varied the concentrations of other signaling proteins in the apoptotic pathway in the same

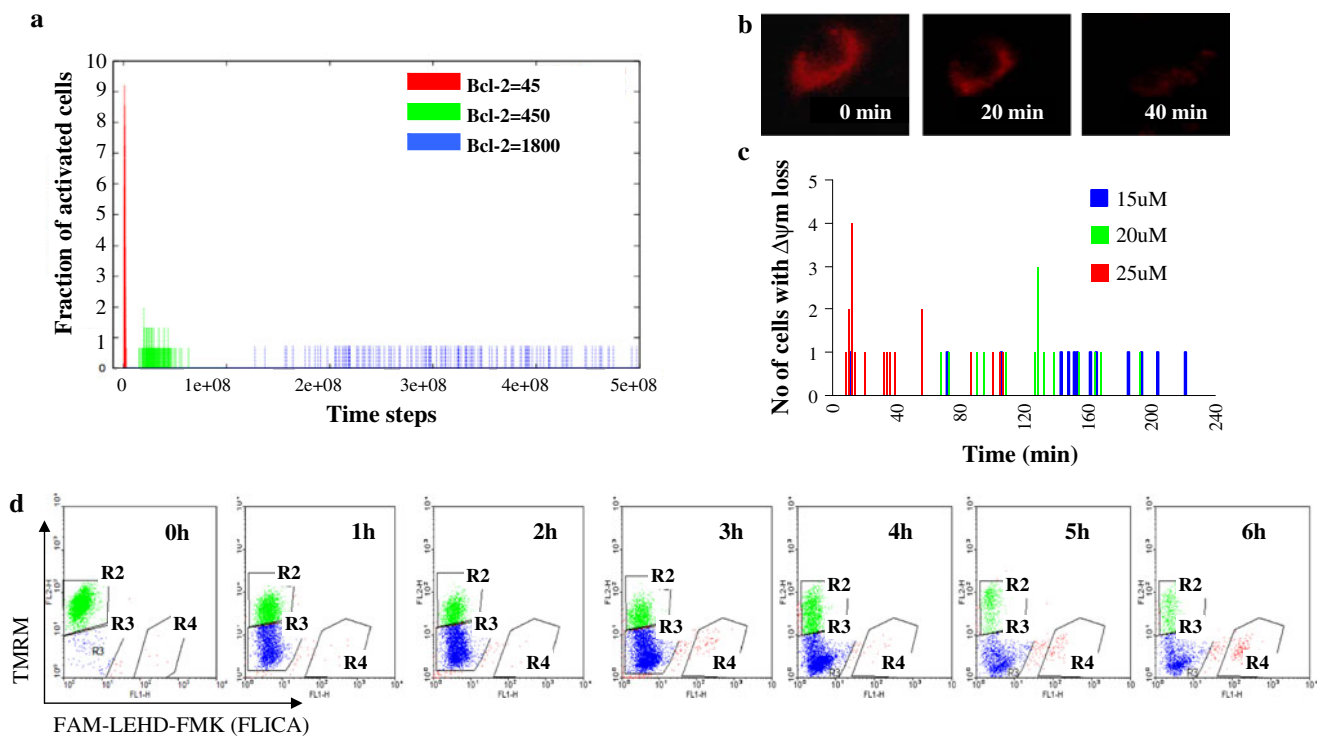

**Fig. 3** Bcl-2 increases cell-to-cell variability in time-to-MOMP, which contributes significantly to cell-to-cell variability in time-to-death. **a** The computational model of the intrinsic pathway of apoptosis indicates that Bax2 dimer formation occurs with large cell-to-cell variability, which is modulated by the level of Bcl-2. Note that decrease in Bcl-2 level reduces cell-to-cell variability in time to Bax2 dimer formation. Time is measured in Monte-Carlo (MC) steps. 1 MC step $= 10^{-4}$ s. **b** Mitochondrial depolarization indicates MOMP in SH-SY5Y cells and is displayed as a loss in TMRM fluorescence. Time stamps indicate time after HA14-1 addition. **c** Non-cumulative frequency distribution for time to initial loss of $\Delta\psi_m$ in SH-SY5Y cells ($n \geq 16$ per treatment), based on changes in pixel intensity in the TMRM-sensitive channel. HA14-1-treated HEK293 cells, which are not Bcl-2-dependent, did not respond with the loss of TMRM fluorescence under the same imaging conditions (not shown). **d** Jurkat T cells were treated with HA14-1 (10 μM) for the time indicated, labeled with FLICA, TMRM and 7-AAD, and analyzed by flow cytometry. Cells with permeabilized plasma membrane (7-AAD positive) were excluded from analysis. Region R2, cells with energized mitochondria and lack of caspase 9 activation; region R3, cells with loss of $\Delta\psi_m$ and no or initial caspase 9 activation; region R4, cells with loss of $\Delta\psi_m$ and full caspase 9 activation. Note that (i) cell population is losing TMRM fluorescence gradually, rather then in all-or-none manner; (ii) there is a time delay between the initial loss of TMRM fluorescence and caspase 9 activation; (iii) initial caspase 9 activation occurs in cells with minimal TMRM fluorescence. The number of intermediate events between regions R3 and R4 was less then 5%

manner. The effect of *subtle* variations in protein levels was to increase the cell-to-cell variability in activation of caspase 9 and 3 (Fig. 4, Fig. S7). However, even when all intra- and extracellular parameters are equal between the cells, including lack of variations in protein concentration, significant cell-to-cell variability in $T_d$ was observed in response to the BH3 mimetic (Fig. 4, Fig. S7), leading to fractional cell killing within the time scale of our analysis. We thus show that commitment to cell death is largely impacted due to differences in protein levels at the time of stimulation (Figs. 1b, 4, Fig. S7), which is in agreement with previous studies [10]. Commitment to cell death is also subject to inherent stochastic variations in the molecular signaling reactions, the impact of which depends on the concentration of Bcl-2 (Fig. 1b, Fig. S2) or the dose of BH3 mimetic (Fig. 4, Fig. S7), and may cause decorrelation in $T_d$ in cycloheximide-treated sister cells [10, 20, 32]. Notably, such inherent stochastic variability in apoptotic signaling reactions exists even in the presence of a large number of molecules of each protein species at the time of treatment. Inhibitory signaling reactions and complex network structures dynamically generate a few copy numbers of signaling molecules (such as Bax2 dimers). Moreover, some signaling reactions, such as apoptosome formation, can occur with very low probability. Both factors could contribute to generation of noise in apoptotic signaling.

## Discussion

Other studies have used mathematical models of apoptosis based on ordinary differential equations [e.g., 10], whereas this paper uses probabilistic computational modeling that allows studies of cell-to-cell variability within a given cell

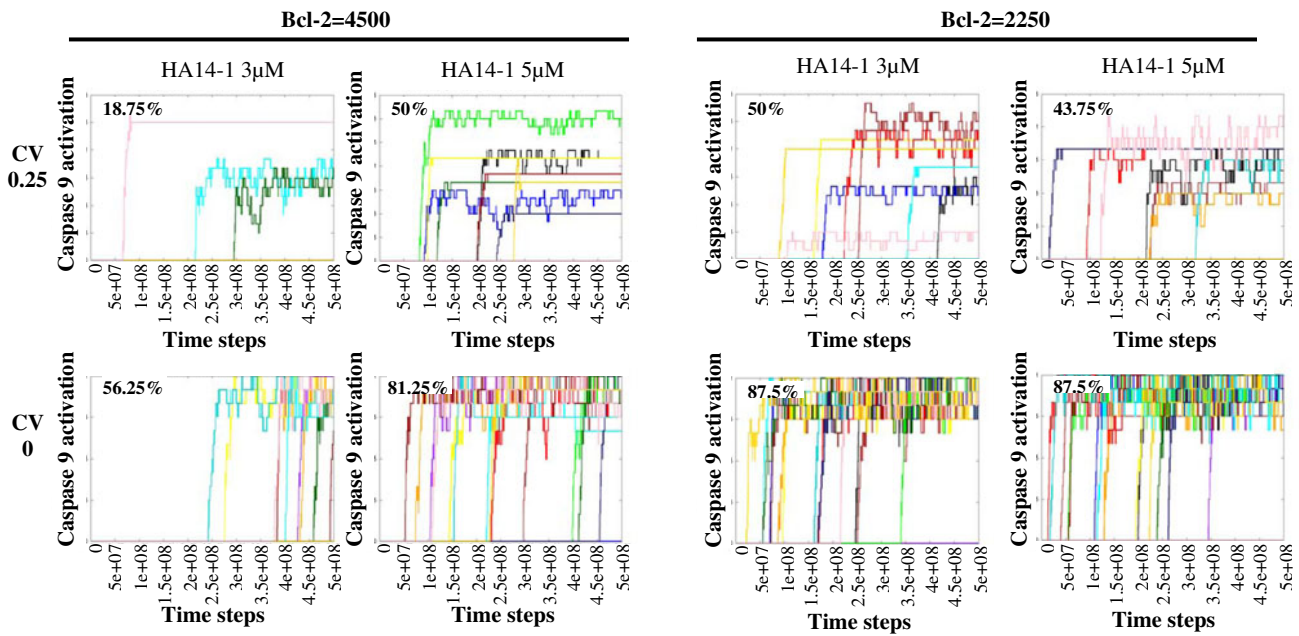

**Fig. 4** Intrinsic stochastic variability in apoptotic signaling is sufficient for cell-to-cell variability in time-to-death. Simulated timing of caspase 9 activation in HA14-1-treated cells. Mean protein concentration of Bcl-2 was set to **a** 4500 molecules ($\sim$7.5 μM), or **b** 2250 molecules ($\sim$3.75 μM). Concentrations of all proteins in the model vary with coefficient of variation (CV) of 0.25 (*upper panels*) or 0 (*lower panels*). Each *line* represents a single simulated cell. *Insets* indicate % cell death within the time of analysis. Total of 16 cells were analyzed. Time is measured in Monte-Carlo (MC) steps. 1 MC step = $10^{-4}$ s

line. We verify qualitatively our model of the intrinsic pathway of apoptosis by using single-cell experimental approach. To model the varying level of Bcl-2 protein experimentally, we use BH3 mimetic HA14-1, which has been previously shown to trigger the mitochondrial pathway of apoptosis via the Bcl-2-Bax axis (26). Although we can not exclude the possibility of off-target effects of HA14-1, such as binding to other Bcl-2 family proteins, or generation of ROS, which is associated not only with degradation of small molecule BH3 mimetics but also with mitochondrial stress, we use this system to provide general qualitative verification of the computational model. Our observations show that as the doses of BH3 mimetic increase, the cell-to-cell variability in time to death, and time to initial loss of mitochondrial potential, decrease. The bi-stability in caspase activation is also experimentally observed in cells treated with HA14-1. This is consistent with the results of the computational model of the intrinsic pathway of apoptosis, explaining how varying levels of Bcl-2 regulate cell-to-cell variability in time-to-death ($T_d$). The model reveals that the time-scale for Bax activation as well as its cell-to-cell variability is significantly increased as the Bcl-2 level increases, especially beyond a threshold Bcl-2 level ($\sim$0.75 μM). Bax activation, in turn, modulates mitochondrial events in such a manner that $T_d$, as measured by caspase 9 and caspase 3 activation, and the cell-to-cell variability in commitment to apoptosis preserve the effect of large Bcl-2 variations. With the current experimental approach we confirm qualitatively the findings of the computational model. In the future more specific interfering with the function of Bcl-2, possibly with the use of more stable and specific next generation BH3 mimetics, could help to achieve better quantitative agreement with the model. Nevertheless, based on the present findings we postulate that as cells with high Bcl-2 levels submit to cell death with large cell-to-cell stochastic variability, a cell with a particularly long $T_d$ can escape apoptosis. Such a long time to death may lead to single cells escaping apoptosis, allow time to acquire other tumor initiating features and finally clonal initiation of tumor growth.

Using the experimentally-verified probabilistic model of the intrinsic pathway of apoptosis we finally study the phenomenon of fractional cancer cell killing which has been widely observed both in vitro and in vivo. Treatments that fail to accomplish complete eradication of cancer cells lead to relapse and often an accelerated tumor growth that exceeds that of the original tumor, particularly if surviving cells have tumorigenic capacity [33, 34]. This is likely to occur especially in cancers that have very common tumor-initiating cells [35]. The heterogeneity in response of cancer cells is often attributed to subtle differences intrinsic to cancer cells or their microenvironment, variation in the access of tumor cells to a drug, or drug insensitivity during certain phases of cell cycle [10, 36]. Although these factors can clearly influence the sensitivity of cancer cells to chemotherapy, as we also show for varying concentrations of Bcl-2, we ask here what would be the outcome of BH3 mimetic treatment in a hypothetical population of identical cancer cells, the behavior of which is not influenced by epigenetic or genetic variations, de novo protein synthesis, induction of survival pathways, or subtle variations in protein concentration. This is achieved by maintaining constant all parameters, but the dose of HA14-1. We hypothesize that the very nature of apoptotic signaling is such that single cells will be able to escape death. We show that treatment of such a population of cancer cells still leads to fractional cell killing which is attributed entirely to the stochastic nature of apoptotic signaling, despite the presence of a large number of molecules for each protein species in the intrinsic pathway of apoptosis [17–19]. Our observation challenges the widely appreciated concept of anti-cancer treatments targeted to induce apoptosis. High levels of Bcl-2 in cancer cells increase even further the inherent cell-to-cell variations in cell fate, increasing the likelihood of fractional cell killing. These observations imply that a very high concentration of BH3 mimetic, most likely incompatible with the survival of non-cancerous cells, would be required to achieve certainty of killing the whole population of cancer cells. Considering the inherent cell-to-cell variability in the apoptotic pathway, approaches that widen the therapeutic window of pro-apoptotic agents (e.g., specific cancer-targeted delivery with bacterially derived minicells [37], combinatorial treatments with drugs that delay mitotic exit [38], or even with standard chemotherapeutic agents [14, 39]), and thus allow significantly increasing the local drug concentration/efficacy, could provide better outcomes. Otherwise, the strategy of adaptive therapy [40], targeting cancer cell cycle or inducing cancer cell senescence may be more effective than treatment protocols aiming at killing all tumor cells.

**Acknowledgments** We are grateful to A. Turner (University of Auckland) for his excellent assistance with time-lapse microscopy, and to Dr D. Wlodkowic (University of Glasgow) for advice on live-cell imaging. We also acknowledge M. Djendjinovic for help with the figure preparation and K. Katipally for computational work during the initial phase of this study.